\documentclass[12pt]{article}
\usepackage{amsmath}
\usepackage{epsf}
\usepackage{epsfig}
\usepackage{here}
\usepackage{amssymb}
\usepackage{citesort}
\usepackage{graphicx}
\usepackage{latexsym}
\newcommand{\be}{\begin{equation}}
\newcommand{\ee}{\end{equation}}
\newcommand{\bear}{\begin{eqnarray}}
\newcommand{\ear}{\end{eqnarray}}

\newsavebox{\LSIM}
\sbox{\LSIM}{\raisebox{-1ex}{$\ \stackrel{\textstyle<}{\sim}\ $}}
\newcommand{\lsim}{\usebox{\LSIM}}
\newsavebox{\GSIM}
\sbox{\GSIM}{\raisebox{-1ex}{$\ \stackrel{\textstyle>}{\sim}\ $}}
\newcommand{\gsim}{\usebox{\GSIM}}

\begin{document}
\begin{titlepage}
\begin{flushright}
BA-03-11\\
DESY 03-110\\
hep-ph/0309252
\end{flushright}
$\mbox{ }$
\vspace{.1cm}
\begin{center}
\vspace{.5cm}
{\bf\Large Seesaw Mechanism in Warped Geometry}\\[.3cm]
\vspace{1cm}
Stephan J. Huber$^{a,}$\footnote{stephan.huber@desy.de}
and 
Qaisar Shafi$^{b,}$\footnote{shafi@bartol.udel.edu} \\ 
 
\vspace{1cm} {\em  
$^a$Deutsches Elektronen-Synchrotron DESY, Hamburg, Germany}\\[.2cm] 
{\em $^b$Bartol Research Institute, University of Delaware, Newark, USA} 

\end{center}
\bigskip\noindent
\vspace{1.cm}
\begin{abstract}
We show how the seesaw mechanism for neutrino masses can be realized
within a five dimensional (5D) warped geometry framework. Intermediate
scale standard model (SM) singlet neutrino masses, needed to explain the
atmospheric and solar neutrino oscillations, are shown to be proportional
to $M_{\rm Pl}\exp((2c-1) \pi kR)$, where $c$ denotes the coefficient of the 5D Dirac
mass term for the singlet neutrino which also has a Planck scale Majorana mass
localized on the Planck-brane, and $kR\approx 11$ in order to resolve the gauge
hierarchy problem. The case with a bulk 5D Majorana mass term for the singlet
neutrino is briefly discussed.
\end{abstract}
\end{titlepage}
\section{Introduction}
A particularly intriguing resolution of the gauge hierarchy problem
is provided by a setting based on five dimensional (5D) warped geometry 
\cite{RS} (see also \cite{G}).
Without invoking supersymmetry it is possible to derive the 'low energy'
TeV scale from 5D Planck scale quantities. Indeed, it may even be possible
to derive even smaller scales, such as TeV$^2/ M_{\rm Pl} \sim 10^{-3}$ eV 
\cite{HSCC} which
may shed some light in the understanding of the observed vacuum energy
density. The warped framework has some other interesting features.
It sheds new light on fermion mass hierarchies and mixings \cite{GP,HS2,H},
and also allows one to accommodate the observed solar and atmospheric neutrino 
oscillations through dimension five operators, without invoking any additional
fields beyond those present in the SM \cite{HS4}.
By introducing SM singlet fermions the observed neutrinos can turn into
light Dirac particles \cite{GN,HS3}.
The approach seems to be consistent with the attractive idea of grand 
unification \cite{GUT}.
Last but by no means least, this approach can be experimentally tested, 
hopefully at the LHC. In particular, the first KK excitations of the SM 
particles are expected to lie in the multi (7-10) TeV range 
\cite{HS,HLS,CET,HPR}. In the presence of brane-localized kinetic 
terms the KK scale may be somewhat lower \cite{branekin}.
A left-right symmetric gauge group in the bulk may also 
bring down the KK scale to a few TeV \cite{LR}.

In this paper we investigate how the four dimensional seesaw mechanism
can be incorporated within the warped setting. This means that one should
understand how an intermediate mass scale for the SM singlet neutrinos arises,
starting with Planck scale quantities. We show how this works out by introducing
in particular 5D Dirac masses for the SM singlet fields, in addition to the 
Majorana masses. 
It remains to be seen if the appearance of an intermediate mass scale for 
'right handed' neutrinos 
can be exploited to yield not only the required light neutrino masses but to also 
realize the observed baryon asymmetry via leptogenesis \cite{FY}. 

\section{KK reduction with a Majorana mass term}
We take the fifth dimension to be an $S_1/Z_2$ orbifold with 
a negative bulk cosmological constant, bordered by two 3-branes 
with opposite tensions and separated by distance $R$.  Einstein's
equations yield the non-factorizable metric  \cite{RS}
\begin{equation} \label{met}
ds^2=e^{-2\sigma(y)}\eta_{\mu\nu}dx^{\mu}dx^{\nu}-dy^2, ~~~~\sigma(y)=k|y|
\end{equation}
which describes a slice of AdS$_5$. The 4-dimensional metric is 
$\eta_{\mu\nu}$=${\rm diag}(1,-1,-1,-1)$, $k$ is the AdS curvature 
related to the bulk cosmological constant and brane tensions, 
and $y$ denotes the fifth coordinate. The AdS curvature and the 
5D Planck mass $M_5$ are both assumed to be of order $M_{\rm Pl}=1.2\times10^{19}$ GeV.
The AdS warp factor $e^{-\pi k y}$
generates an exponential hierarchy of energy scales.
If the brane separation is $kR\simeq 11$, the natural scale at
the negative tension brane, located at $y=\pi R$, is of TeV-size, 
while the scale at the brane at $y=0$ is of order $M_{\rm Pl}$. 

We consider the fermionic action on the warped background (\ref{met})
\begin{equation} \label{5Daction}
S=\int d^4x\int_{-\pi R}^{\pi R} dy\sqrt{-G}\left(\bar \Psi iE_a^M\gamma^a
(\partial_M+\omega_M)\Psi-m_D \bar \Psi \Psi -m_M \bar \Psi \Psi^c\right),
\end{equation}
where $E^M_a$ is the f\"unfbein and $\gamma^a=(\gamma^{\mu},\gamma^5)$
represent the Dirac matrices in flat space. The index $M$ refers to objects
in curved 5D space, the index $a$ to those in tangent space.
The spin connection related to the metric (\ref{met}) is found to be 
$\omega_M=(\frac{1}{2}\sigma'e^{-\sigma}\gamma^5\gamma_{\mu},0)$, with
$\sigma'=d\sigma/d y$.
$\Psi^c=C_5\gamma^0\Psi^*$ is the charge conjugated
spinor.

Fermions in 5D are non-chiral.
Chirality in the 4D low energy effective theory is restored by 
the orbifold boundary conditions. Fermions have two possible 
transformation properties under the $Z_2$ orbifold symmetry,
$\Psi(-y)_{\pm}=\pm i\gamma_5 \Psi(y)_{\pm}$, depending on
whether the left- or right-handed components are chosen to be even.
Thus, $\bar\Psi_{\pm}\Psi_{\pm}$ 
is odd under $Z_2$, and the Dirac mass parameter, which is also odd, 
can be parametrized as $m_D=-c\sigma'$.\footnote{The minus
sign in the definition ensures that the meaning of $c$ matches
with refs.~\cite{GP,HS2}, which use a different signature of the metric.}  
The bilinear $\bar\Psi_{\pm}\Psi_{\pm}^c$ is even, resulting in
an even Majorana mass $m_M$.  The Majorana mass can
have bulk and boundary contributions. 
The boundary mass terms are restricted only by 4D Lorentz invariance
and one could think of choosing them differently for the left- and right-handed 
components of the Dirac spinor. However, boundary mass terms
are only felt by the even components. The odd components do
have only derivative couplings to the boundary. 

In the following we perform the KK reduction of the action (\ref{5Daction}) 
to four dimensions. Without the Majorana mass $m_M$ this has 
first been discussed in ref.~\cite{GN} (see also \cite{GP}). 
Using the warped metric (\ref{met}) and defining 
$\hat\Psi=e^{-2\sigma}\Psi$ we obtain
\begin{equation} \label{5Daction1}
S=\int d^4x\int_{-\pi R}^{\pi R} dy\left(\bar{\hat{\Psi}} (ie^{\sigma}\gamma^{\mu}
\partial_{\mu}+i\gamma^5\partial_5)\hat\Psi-m_D \bar{\hat{\Psi}} \hat\Psi -
m_M \bar{\hat{\Psi}} \hat\Psi^c\right).
\end{equation}
We decompose the 5D fields as 
\begin{equation}
\hat\Psi_{L,R}(x^{\mu},y)=\frac{1}{\sqrt{2\pi R}}\sum_{n=0}^{\infty}\Psi^{(n)}_{L,R}
(x^{\mu})f_{L,R,n}(y),
\end{equation}
where $\Psi_{L,R}=\pm i\gamma^5\Psi_{L,R}$. For non-vanishing $m_M$
the spectrum of KK states is no longer vector-like. Instead, it consists of an 
infinite tower of Majorana fermions with masses $m_n$. 
Requiring that after $y$ integration 
the action (\ref{5Daction1}) 
reduces to the usual action of massive Majorana fermions in four dimensions,
the wave functions $f_{L,R,n}$ must obey the conditions
\begin{eqnarray} \label{eq1}
-m_Mf_{L,n}-(\partial_5+m_D)f^*_{R,n}&=&-m_ne^{\sigma}f^*_{L,n}\nonumber\\
-m_Mf_{R,n}^*+(\partial_5-m_D)f_{L,n}&=&-m_ne^{\sigma}f_{R,n}.
\end{eqnarray} 
To arrive at these expressions we have used the 4D Majorana condition 
$\bar\Psi_R^{(n)}=\Psi_L^{(n)}$. For $m_M=0$ we reproduce
the results of ref.~\cite{GN}. The normalization conditions read
\begin{equation}
\frac{1}{2\pi R}\int^{\pi R}_{-\pi R}dye^{\sigma}(f^*_{L,m}f_{L,n}+f^*_{R,m}f_{R,n})
=\delta_{mn}.
\end{equation}
Notice that for non-vanishing Majorana mass $f_{L,n}$ and $f_{R,n}$ are 
no longer complete sets of functions by there own. 
If $m_M$ is real,
the eqs.~(\ref{eq1}) can be split into real parts
\begin{eqnarray} \label{real}
-m_M{\rm Re}f_{L,n}-(\partial_5+m_D){\rm Re}f_{R,n}&=&
-m_ne^{\sigma}{\rm Re}f_{L,n}\nonumber\\
-m_M{\rm Re}f_{R,n}+(\partial_5-m_D){\rm Re}f_{L,n}&=&-m_ne^{\sigma}{\rm Re}f_{R,n}
\end{eqnarray} 
and imaginary parts
\begin{eqnarray} \label{im}
-m_M{\rm Im}f_{L,n}+(\partial_5+m_D){\rm Im}f_{R,n}&=&
m_ne^{\sigma}{\rm Im}f_{L,n}\nonumber\\
m_M{\rm Im}f_{R,n}+(\partial_5-m_D){\rm Im}f_{L,n}&=&-m_ne^{\sigma}{\rm Im}f_{R,n}.
\end{eqnarray} 
The eqs.~(\ref{real}) and (\ref{im})
are related by $m_M\rightarrow-m_M$. For a complex Majorana
mass eqs.~(\ref{real}) and (\ref{im}) no longer separate.

For $m_M=0$ eqs.~(\ref{real}) and (\ref{im}) allow for
a chiral zero mode solution, and the chirality depends on the
chosen orbifold boundary conditions. If the Majorana mass term 
is turned on, the zero mode picks up a mass and becomes
a mixture of left- and right-handed states. We still can decouple
left- and right-handed states in eqs.~(\ref{real}) and (\ref{im})
and end up, for instance, with
\begin{equation}
-m_M{\rm Re}f_{R,n}-(\partial_5-m_D)\frac{1}{m_ne^{\sigma}-m_M}
(\partial_5+m_D){\rm Re}f_{R,n}=-m_ne^{\sigma}{\rm Re}f_{R,n}.
\end{equation}
This equation is complicated but can be solved numerically. Taking
into account the boundary conditions, 
e.g.~${\rm Re}f_{R,n}(0)={\rm Re}f_{R,n}(\pi R)=0$ 
for odd right-handed modes, the
spectrum of KK  masses can be determined. Potential problems
arise if $1/(m_ne^{\sigma}-m_M)$ becomes singular.

A particularly simple case arises if the Majorana mass is
confined to a boundary. Then we can build the wave 
functions from the $m_M=0$ solutions \cite{GN,GP}
\begin{eqnarray}
{\rm Re}f_{L,n}(y)=\frac{e^{\sigma/2}}{N_n}\left[ J_{-c-1/2}(\frac{m_n}{k}e^{\sigma})
+b(m_n)Y_{-c-1/2}(\frac{m_n}{k}e^{\sigma}) \right] \nonumber \\
{\rm Re}f_{R,n}(y)=\frac{e^{\sigma/2}}{N_n}\left[ J_{-c+1/2}(\frac{m_n}{k}e^{\sigma})
+b(m_n)Y_{-c+1/2}(\frac{m_n}{k}e^{\sigma}) \right],
\end{eqnarray}
with $b(m_n)=-J_{-c+1/2}(\frac{m_n}{k}\Omega)/Y_{-c+1/2}(\frac{m_n}{k}\Omega)$.
The warp factor is defined as $\Omega=e^{\pi kR}$.
The Majorana mass shows up only in the boundary conditions.
If the Majorana mass is confined to the Planck-brane, i.e. $m_M=d\cdot \delta(y)$,
we find
\begin{eqnarray} \label{BC}
{\rm Re}f_{R,n}(0)-\frac{d}{2}{\rm Re}f_{L,n}(0)&=&0 \nonumber \\
{\rm Re}f_{R,n}(\pi R)&=&0,
\end{eqnarray}
where we have chosen $f_R$ to be odd.
These equations demonstrate the coupling between left- and
right-handed states which is introduced by the Majorana mass term.
Taking $d=0$ we recover the result of \cite{GN}. The spectrum
of KK masses $x_n=m_n/k$ is finally obtained from
\begin{equation}
\left|\begin{array}{cc} 
J_{-c+1/2}(x_n\Omega) & Y_{-c+1/2}(x_n\Omega) \\[.1cm] 
J_{-c+1/2}(x_n)-\frac{d}{2}J_{-c-1/2}(x_n) & 
Y_{-c+1/2}(x_n)-\frac{d}{2}Y_{-c-1/2}(x_n)
\end{array}\right|=0.
\end{equation}  
The analogous expressions for the imaginary parts of $f_{L,R}$ 
are obtained by switching the sign of the Majorana mass.

\section{The KK spectrum}
For a vanishing Majorana mass the KK spectrum of a bulk
fermion consists of a chiral zero mode, which we choose to be
left-handed, and a tower of
excited vector-like states. The location of the zero mode
depends on the bulk Dirac mass \cite{GN}
\begin{equation}
f_{L,0}(y)=\frac{e^{-ck|y|}}{N_0}.
\end{equation} 
For $c>1/2$ $(c<1/2)$ the zero mode is localized 
near the boundary at $y=0$ $(y=\pi R)$, i.e.~at the Planck- (TeV-) brane.
The excited KK states are always localized at the TeV-brane. 

If we turn on a small Majorana mass, the zero mode
picks up a mass. Its wave function receives a non-vanishing
odd (right-handed) component. Defining
\begin{equation}
r_{L,R,n}=\frac{1}{2\pi R}\int^{\pi R}_{-\pi R}dye^{\sigma}f^*_{L,R,n}f_{L,R,n},
\end{equation}
the even content of the wave function $f_n=(f_{L,n},f_{R,n})$ is given by
\begin{equation} \label{comp}
r_{n}=\frac{r_{L,n}}{r_{L,n}+r_{R,n}}.
\end{equation}
\begin{figure}[t] 
\begin{picture}(100,170)
\put(5,10){\epsfxsize7cm \epsffile{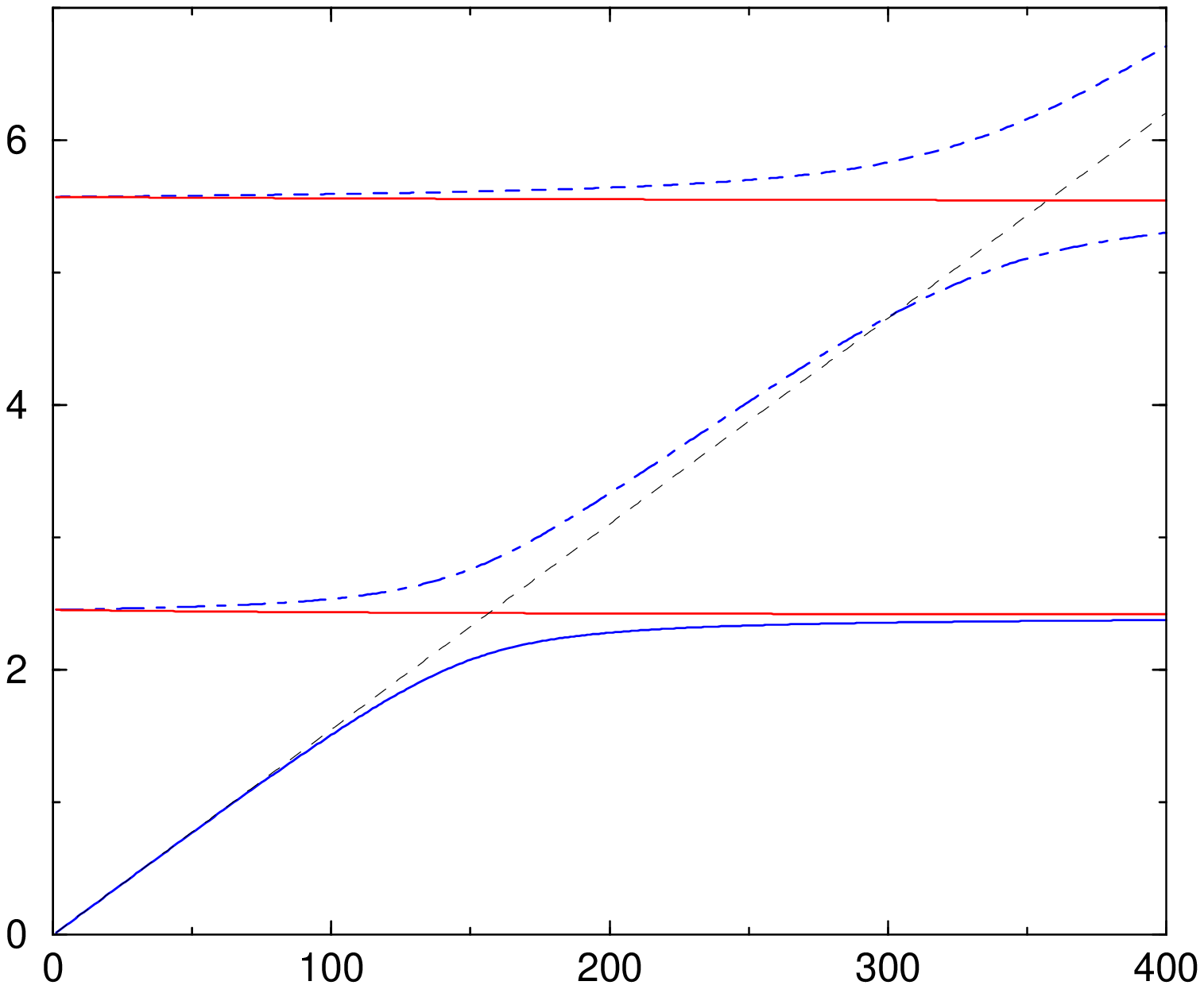}}
\put(220,10){\epsfxsize7.3cm \epsffile{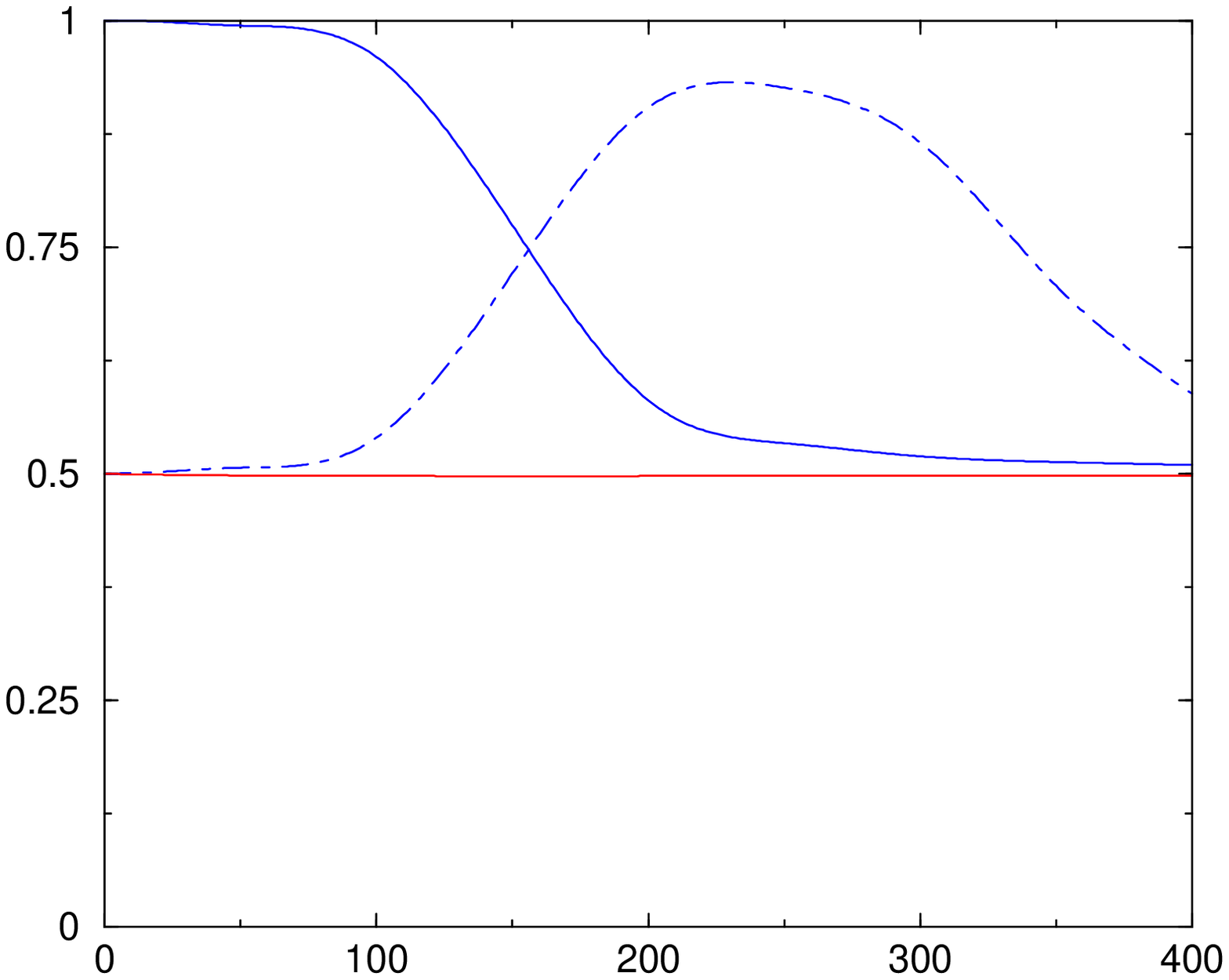}}
\put(20,160){(a)}
\put(400,160){(b)}
\put(40,30){\small $0_+$}
\put(100,100){\small $1_+$}
\put(160,158){\small$2_+$}
\put(30,60){\small$1_-$}
\put(30,130){\small$2_-$}
\put(280,150){\small $0_+$}
\put(370,140){\small $1_+$}
\put(310,80){\small$1_-$}
\put(-20,130){{$x_n\Omega\uparrow$}} 
\put(205,150){{$r_{n}\uparrow$}} 
\put(160,-5){$d\Omega\longrightarrow$}
\put(380,-5){$d\Omega\longrightarrow$}
\end{picture} 
\caption{The KK masses (a) and the even content (b) of the lowest KK
states as a function of the Majorana mass $m_M=d\cdot \delta(y)$. We 
have taken $c=1/2$ and $\Omega=10^{14}$.
}
\label{f_1}
\end{figure}
The vector-like pairs of excited states split up. Once the Majorana mass becomes
larger than a critical value, the zero mode reaches the KK scale and disappears 
from the low energy spectrum.

Let us discuss the case of a Majorana mass confined to the 
Planck-brane in more detail. As long as $d\lsim \Omega^{-2c}$ 
there exists an (almost) chiral mode with mass
\begin{eqnarray} \label{chiral}
x_0\Omega&\approx& \frac{d}{2}(1-2c)\Omega^{2c}, \quad c\lsim \frac{1}{2}\nonumber \\
x_0\Omega&\approx& \frac{d}{2}(2c-1)\Omega, \quad c\gsim \frac{1}{2}\nonumber \\
x_0\Omega&\approx& 0.015\cdot d, \quad c=\frac{1}{2}.
\end{eqnarray}
The splitting of the masses of the excited states is proportional to $d/x_n$. 
Their overall mass is almost unchanged.
This behavior becomes clear from fig.~\ref{f_1}a, where  
we present the lowest KK masses as a function of the Majorana mass $d$.
In this example we have taken the parameters $c=1/2$ and $\Omega=10^{14}$.
We have labeled the states $i_{\pm}$ depending on whether they arise from 
eqs.~(\ref{real}) or (\ref{im}). For $d>0$ the ``zero mode'' belongs to eq.~(\ref{im}).
At  $d\Omega\approx150$ the mass of $0_+$ becomes comparable to the first
KK mass, where it saturates, while the mass of $1_+$ starts to increase.
If the Majorana mass is further increased this phenomenon happens at 
higher KK levels, i.e.~the states $n_+$ join the KK level $(n+1)$. During this
process the masses of the states $n_-$ remain practically constant.
For large Majorana masses, in our example $d\Omega\gg 150$ the
mass splitting in the KK level formed by $n_+$ and $(n+1)_-$
s proportional to $x_n/d$.

It is instructive to study the content of even states among the wave
functions, which we present in  fig.~\ref{f_1}b. For $d=0$ we have 
$r(0_+)=1$, which means that there is truly a chiral zero mode. The excited
states are perfect even-odd mixtures, i.e.~$r=1/2$. If a
Majorana mass is turned on, $r(n_+)$ changes, while the
content of the $n_-$ states is not significantly changed. 
In the range of $d$ where $x(1_+)$ is rapidly growing, $1_+$
becomes an almost pure even state. This means that as
the Majorana mass is increasing an almost even state (``chiral state'',
of course it is a massive state!)
is moving through the KK spectrum.

\begin{figure}[t] 
\begin{picture}(100,170)
\put(5,10){\epsfxsize7cm \epsffile{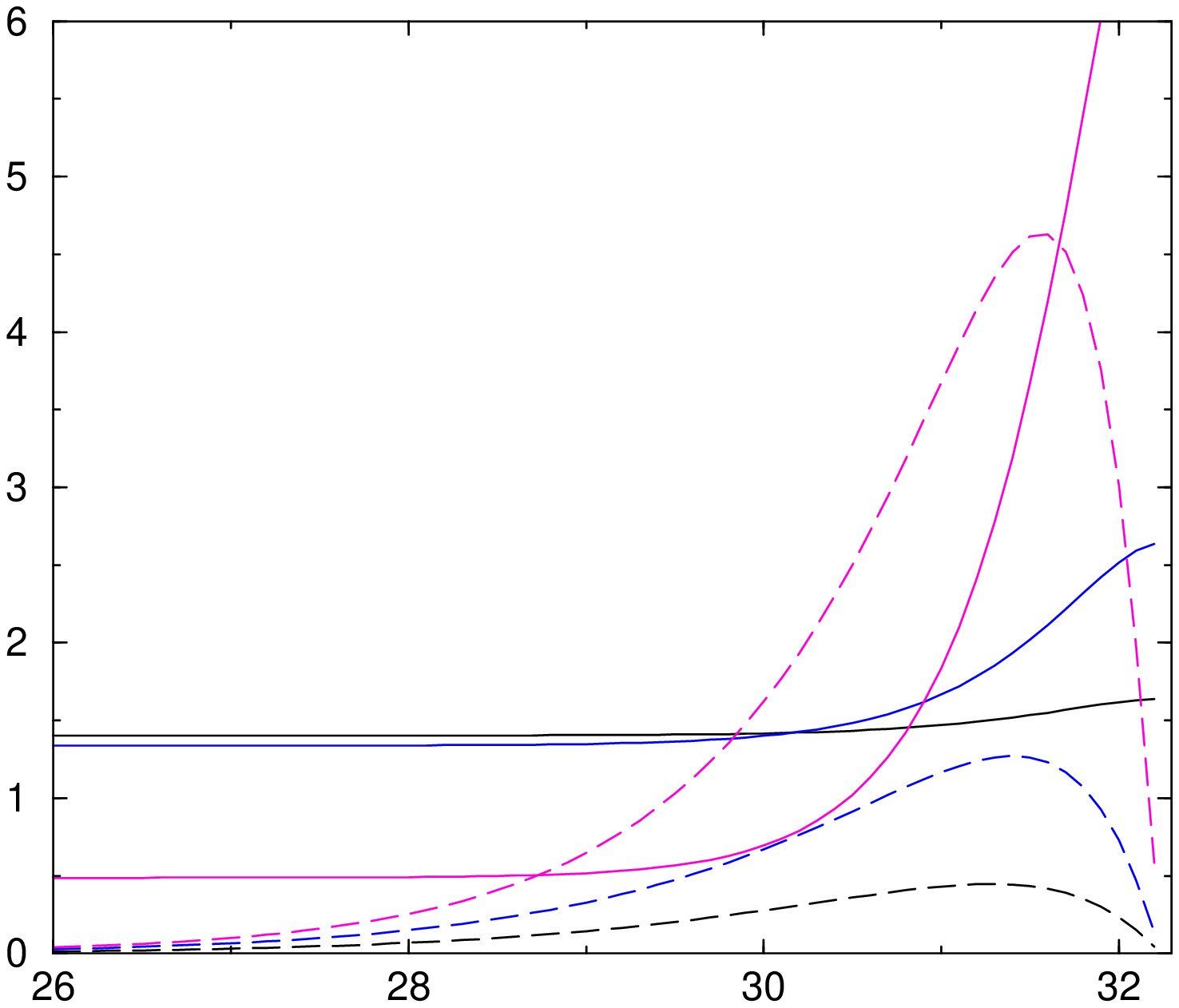}}
\put(220,10){\epsfxsize7.2cm \epsffile{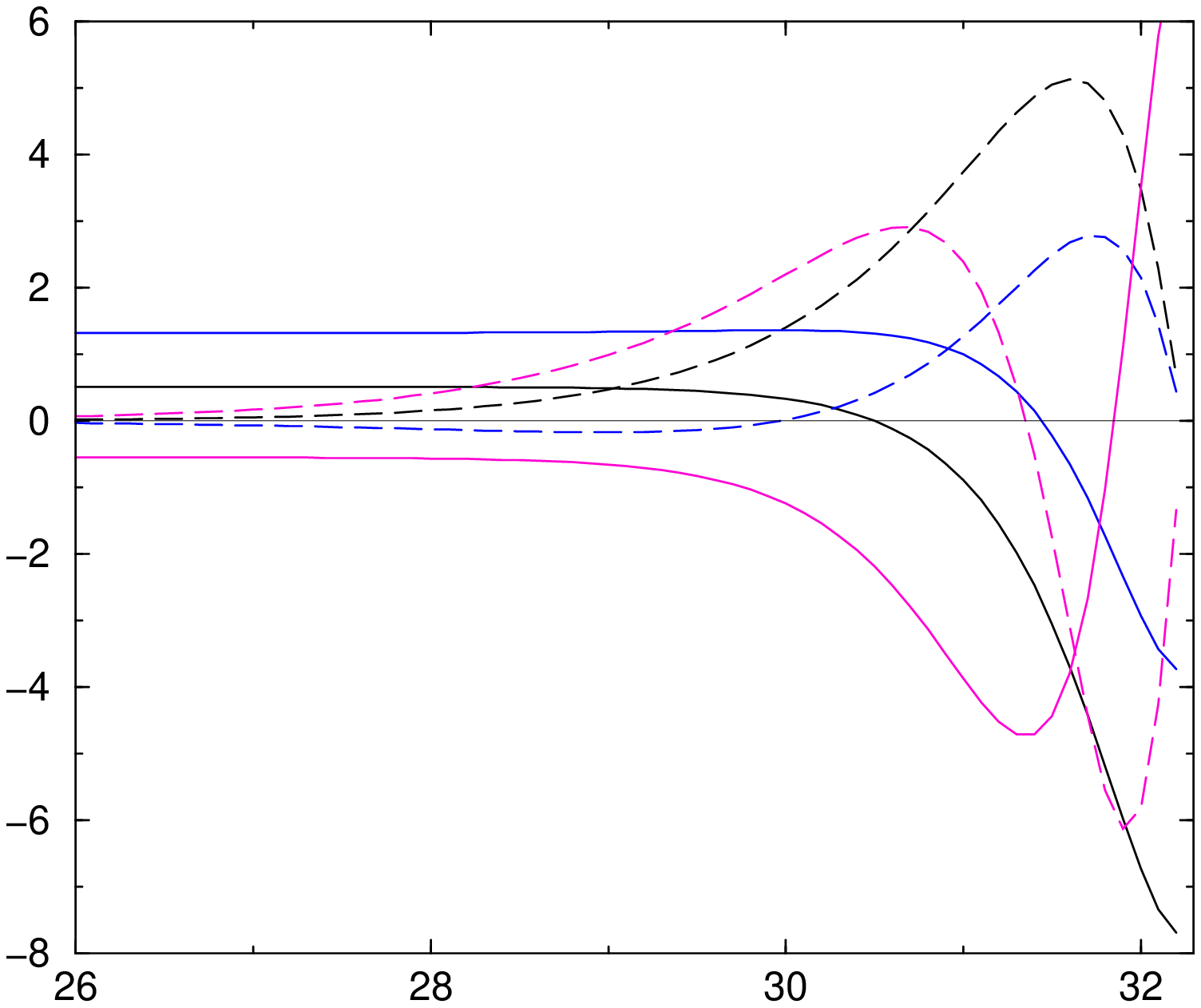}}
\put(25,160){$e^{\sigma/2}f_{0+}$}
\put(245,160){$e^{\sigma/2}f_{1+}$}
\put(168,160){200}
\put(40,44){100}
\put(40,60){50}
\put(143,115){200}
\put(169,41){100}
\put(170,21){50}
\put(270,127){200}
\put(369,91){400}
\put(270,93){100}
\put(340,138){400}
\put(389,143){200}
\put(375,164){100}
\put(160,-3){$yk\longrightarrow$}
\put(380,-3){$yk\longrightarrow$}
\end{picture} 
\caption{The wave functions of the $0_+$ and $1_+$ states in 
the vicinity of the TeV-brane for different values of the 
Majorana mass ($d=50,100,200$ and $d=100,200,400$). 
The left- (right-) handed components are shown in solid 
(dashed) lines. We have taken $c=1/2$ and $\Omega=10^{14}$.
}
\label{f_2}
\end{figure}

In fig.~\ref{f_2} we present the wave functions of the $0_+$ 
and $1_+$ states. The odd component of $0_+$ becomes more 
and more important as we increase $d$ from 50 to 200. At the 
same time the even part of $0_+$ gets suppressed in the bulk 
and gets localized towards the TeV-brane like an excited state. 
The state $1_+$ is localized towards the TeV-brane for $d=100,400$.
For $d=200$ its even component becomes somewhat delocalized.
At the same time the amplitude of the odd component shrinks, as
expected from fig.~\ref{f_1}b. 

Using eqs.~(\ref{real}) and (\ref{im}), the Majorana mass is
included in the KK reduction from the very beginning. This 
procedure is analogous to our treatment of boundary masses
of gauge bosons in refs.~\cite{HS,HLS}. Alternatively, the KK 
reduction can be done with a vanishing Majorana mass. When
the 5D action is integrated over the extra dimension, the
Majorana mass term induces additional operators which mix
the different KK states. The general mass matrix ${\cal M}$ is given by
\begin{equation} \label{mass}
{\cal L}_M= (\Psi_L^{(0)},\Psi_L^{(1)}, \bar\Psi_R^{(1)},\dots) \left(\begin{array}{cccc} 
A_{00} & A_{01} & 0 &\cdots \\[.1cm] 
A_{01} & A_{11} & D_1& \cdots \\[.1cm]  
0 &  D_1 & B_{11} & \cdots \\
\vdots & \vdots & \vdots & \ddots
\end{array}\right)\left(\begin{array}{c}\Psi_L^{(0)} \\[.1cm]   \Psi_L^{(1)} \\[.1cm]   
\bar\Psi_R^{(1)} \\ \vdots \end{array}\right) ,
\end{equation} 
where $D_n$ is the $n$th KK mass and
\begin{eqnarray} \label{AB}
A_{mn}&=&\int_{-\pi R}^{\pi R}\frac{dy}{2\pi R}m_M(y)f_{L,m}(y)f_{L,n}(y)\nonumber \\
B_{mn}&=&\int_{-\pi R}^{\pi R}\frac{dy}{2\pi R}m_M(y)f_{R,m}(y)f_{R,n}(y).
\end{eqnarray}
Note that $f_{L,R,n}$ and $\Psi_{L,R}^{(n)}$ here denote the fields
and wave functions obtained with a vanishing Majorana mass, while
earlier these symbols were used for the mass eigenstates including
the Majorana mass.
The zeros in ${\cal M}$ follow from the orbifold $Z_2$ symmetry.
For a boundary Majorana mass $B_{mn}$ vanishes. 

The advantage of eqs.~(\ref{real}) and (\ref{im}) is that they
diagonalize the infinite dimensional mass matrix ${\cal M}$ in 
a single step. However, it turns out that in many cases simple 
finite truncations
of ${\cal M}$ provide valuable information on the KK spectrum.  
Let us focus again on the case of a Planck-brane Majorana mass, where  
$B_{mn}$ vanishes. For $c\gsim 0.3$
the mass spectrum can by reliably computed up to the $n$th 
KK level by taking into account the states $\Psi^{(0)}$ to $\Psi^{(n)}$.
The results rapidly converge if more KK states are included. 
For small values of the Majorana mass, i.e.~as long as $x_0\ll x_1$ or 
$A_{00}\ll D_1$, one finds for the former zero mode a mass of
\begin{equation} \label{zerosmall}
x_0\approx \frac{A_{00}}{k}.
\end{equation}
The mass splitting of the  $n$th KK level is found to be
\begin{equation}
\Delta x_n\approx \frac{A_{nn}}{k}.
\end{equation}
For large Majorana masses, $A_{00}\gg D_1$, the mass of
the almost even (``chiral'') state, which is moving up the spectrum,
is approximately given by $A_{00}$. In the next section the mass of 
this state will be identified with the seesaw mass scale.
The mass splitting of the 
first KK level reads
\begin{equation} \label{21}
\Delta x_1\approx \frac{A_{11}D_1^2}{(A_{00}+A_{11})^2k}.
\end{equation}
For  $c\lsim 0.3$ the former zero mode becomes closely
localized towards the TeV-brane. Then the $A_{ij}$ are no longer
dominated by $A_{00}$ and eq.~(\ref{21}) receives non-negligible 
corrections from higher KK modes. For a Majorana mass term on the 
TeV-brane  eq.~(\ref{21}) receives corrections as well. 

The truncated mass matrix (\ref{mass}) can be used to
study Majorana mass profiles for which eqs.~(\ref{real}) and (\ref{im})
are not analytically solvable. Let us discuss the case of a
homogeneous bulk Majorana mass $m_M(y)=d\cdot k$. For
$d\ll \Omega^{-1},\Omega^{-2c},1$, where $c\gsim1/2,~ 0\lsim c \lsim1/2, ~c\lsim0$,
there is still a light mode, whose mass
is given by eq.~(\ref{zerosmall}). The mass splittings of the 
KK levels receive corrections from the non-vanishing $B_{mn}$.
For the first KK level one finds $\Delta x_1\approx(A_{11}+B_{11})/k$.
As long as $c\gsim0$,  $B_{mn}$ turns out to be only a tiny correction of
order $B_{11}/A_{11}\sim\Omega^{-2c}$. 
For $d\sim 1$ the mass splitting
becomes comparable to the splitting between different KK levels.
The pairing of KK states is completely gone. Thus bulk and boundary
mass terms predict a rather different KK spectrum for $d\sim 1$.
A very large Majorana mass $d\gg 1$ does not shift the complete
KK spectrum to higher values. The KK masses in this case depends 
in an oscillatory way on $d$. In the case of flat extra dimensions
this behavior was already found in ref.~\cite{DDG}. 

One could ask under what conditions the bulk and
boundary Majorana mass terms could be responsible for the
observed small neutrino masses $m_{\nu}$, once the bulk fermion field
is identified with a SM neutrino. Because of the SM gauge invariance,
the Majorana mass term must arise from an SU(2) triplet (either elementary
or from two doublets). The gauge hierarchy problem requires Higgs
fields and therefore the Majorana mass term to be localized at the TeV-brane. 
We have studied this scenario in ref.~\cite{HS3}, finding that a tuning
of order $10^{-3}$ to $10^{-9}$, depending on $k/M_{\rm Pl}$, is needed 
to generate sub-eV neutrino masses.  The neutrinos should be localized
towards the Planck-brane. 

Could the Majorana mass terms explain an eV-scale mass for a sterile 
neutrino? The bulk mass is
certainly not a convincing possibility, since the small sterile neutrino 
mass has to be put in by hand in the 5D action. 
A natural value for the sterile neutrino mass could be expected to be
comparable to the KK scale (TeV-size).
If the Majorana mass is localized at the Planck-brane, small sterile neutrino
masses can be achieved by localizing the fermion towards the
TeV-brane. From eq.~\ref{chiral} we can read off that for $c\approx -1/2$
sub-eV masses are possible for $d\sim 1$. Since the neutrino is
sterile, such a small value of $c$ is not disfavored by 
electroweak observables \cite{HLS}. If the Majorana
mass is localized on the TeV-brane, small sterile neutrino masses can
be produced by localizing the fermion towards the Planck-brane with 
$c\approx 1$. 

In the next section
we discuss how a Planck-brane Majorana mass assigned to
a ``right-handed'' bulk neutrino leads to a satisfactory seesaw mechanism.
Realistic neutrino masses can be accommodated without introducing
any small numbers.

\section{The seesaw mechanism in warped geometry}
The seesaw mechanism provides a tiny mass for the SM
neutrinos $\nu_L$ by coupling them to heavy right-handed
neutrinos $N$\cite{seesaw}
\begin{equation} \label{ss}
M_{\nu}=\frac{\lambda_N^2 \langle H \rangle^2}{M_N}.
\end{equation}
Here $M_N$ denotes the Majorana mass for the right-handed
neutrinos and $\lambda_N\nu_LN  H$ is neutrino Yukawa interaction.
Taking $M_{\nu}\sim 50$meV (of the order of the
atmospheric neutrino mass splitting $\sqrt{\Delta m_{\rm atm}^2}$ \cite{ATM}), one finds 
$M_N\sim \lambda_N^2\cdot6\times10^{14}$ GeV.
For $0.01\lsim\lambda_N\lsim1$ this points to an intermediate
scale for the right-handed Majorana mass. 

Naively it seems problematic to implement the seesaw
mechanism in a warped extra dimension. We have seen
in the previous section that despite assigning a
Planck-size $(d\sim1)$ Majorana mass to a bulk fermion, its
lowest KK states have a mass of order $k\Omega^{-1}$, which
is in the TeV region. However, the KK mass is (almost) Dirac-like. 
Inserting it into eq.~(\ref{ss}) does not lead to the correct light 
neutrino mass.

In the following we study the coupling of two bulk fermion fields 
$\nu$ and $N$, corresponding to left- and right-handed neutrinos.
The generalization to three generations is straightforward.
Lepton number is broken by the Majorana
mass of $N$, which we assume is localized at the Planck-brane,
i.e.~$m_M(N)=d\cdot \delta(y)$. Both fields may have bulk Dirac masses 
indicated by $c_{\nu}$ and $c_N$. Let us first discuss the situation
along the lines of eq.~(\ref{mass}), which means leaving out the 
Majorana mass (and the Yukawa interaction) in the KK reduction of $N$.
From the KK reduction of the left-handed neutrino field $\nu$
we obtain a left-handed zero mode $\nu_L^{(0)}$, corresponding to
the SM neutrino,  and an infinite
tower of left- and right-handed KK excited states $\nu_L^{(m)}$ and 
 $\nu_R^{(m)}$. The sterile (right-handed) neutrino
decomposes into the right-handed zero mode $N_R^{(0)}$ and the
KK excited states $N_L^{(m)}$ and $N_R^{(m)}$. In the basis
of $(\nu_L^{(0)},\bar N_R^{(0)}, \nu_L^{(1)},\bar \nu_R^{(1)},\bar N_R^{(1)},N_L^{(1)},...)$
the general mass matrix takes the form 
\begin{equation} \label{nmass}
{\cal M}_{\nu}=\left(\begin{array}{ccccccc}
0 &C_{00} & 0 & 0 & C_{01} & 0 &\cdots \\[.1cm] 
C_{00} & A_{00} & C_{10} & 0 & A_{01} & 0& \cdots \\[.1cm]  
0 & C_{10} & 0 & D_{\nu, 1} & C _{11}& 0 & \cdots \\[.1cm] 
0 & 0 & D_{\nu, 1} & 0 & 0 & C_{o,11} & \cdots \\[.1cm]  
C_{01} & A_{01} & C_{11} & 0 & A_{11} & D_{N,1}& \cdots \\[.1cm]  
0 & 0 & 0 & C_{o,11}   &  D_{N,1} & B_{11} & \cdots \\
\vdots & \vdots & \vdots &\vdots & \vdots & \vdots & \ddots
\end{array}\right).
\end{equation} 
Here $D_{\nu,m}$ and $D_{N,m}$ denote the KK masses
of $\nu$ and $N$, respectively. The mass terms $A_{mn}$
and $B_{mn}$ are defined as in eq.~(\ref{AB}). Because we
have taken a boundary Majorana mass, $B_{mn}$ vanishes.
The mass terms
\begin{eqnarray}\label{C}
C_{mn}&=&\int_{-\pi R}^{\pi R}\frac{dy}{2\pi R}\lambda_N^{(5)}H(y)
f_{L,m}^{(\nu)}(y)f_{R,n}^{(N)}(y)\nonumber \\
C_{o,mn}&=&\int_{-\pi R}^{\pi R}\frac{dy}{2\pi R}\lambda_N^{(5)}H(y)
f_{R,m}^{(\nu)}(y)f_{L,n}^{(N)}(y)
\end{eqnarray}
arise from the Yukawa interaction with 5D coupling $\lambda^{(5)}_N$ 
after electroweak symmetry
breaking. We take the Higgs profile to be strictly confined
to the TeV-brane so that $C_{o,mn}$ vanish. Lepton number is
violated only by the entries $A_{mn}$ and $B_{mn}$.

We can compute the light neutrino mass by truncating the 
mass matrix (\ref{nmass}). Taking more and
more KK states into account, it can be checked numerically that the
procedure indeed converges. In first approximation 
the light neutrino mass is found to be
\begin{equation} \label{mnu}
m_{\nu}\approx\frac{C_{00}^2}{A_{00}}
\left(1-\frac{C_{00}^2A_{11}+C_{01}^2A_{00}-2C_{00}C_{01}A_{01}}
{D_{N,1}^2A_{00}}+\dots\right).
\end{equation}
The first term of this result is completely analogous to the ordinary seesaw
formula (\ref{ss}). The seesaw scale turns out to be 
$A_{00}$, the mass of the heavy ``chiral'' mode in the
spectrum of $N$, which was discussed in the previous 
section.  The relevant Dirac mass in the numerator
arises from the two zero modes. The KK masses of the
excited states do not show up in the leading term since they 
are Dirac-like. They appear as corrections of order ${\cal O}(C^2/D^2)$
in eq.~(\ref{mnu}). The mass terms from the electroweak
symmetry breaking $C_{ij}$ are in the same range as the
charged lepton masses, while the KK scale is TeV-size.
We thus are left with tiny corrections to the 
seesaw formula of oder $10^{-6}$.
A related version of
a warped seesaw mechanism was recently discussed in ref.~\cite{GP2},
where the Higgs field was identified with a slepton
in a (partly) supersymmetric setup. 

The system can of course also be analyzed in the basis
where the Majorana mass is included in the KK decomposition.
\footnote{One could even include in the KK reduction the masses 
from electroweak symmetry breaking as well.}
The disadvantage of this procedure is that the states of the
KK tower of $N$ are no longer strictly Dirac-like and contribute
to the light neutrino mass. Therefore one has to sum up
all contributions up to the heavy ``chiral'' state in the spectrum.
Depending on the size of the Majorana mass and the fermion
locations (i.e. $c$ parameters), the number of relevant states
can be up to order $\Omega$.

\section{Discussion}
The warped seesaw mechanism we just described generates
sub-eV Majorana masses for the SM neutrinos. However,
it is not their only source. In ref.~\cite{HS3} we discussed
neutrino masses from the dimension-5 interaction $(1/M)HHLL$.
Here we assume that this contribution is negligible due to
a small coefficient multiplying the dimension-5 operator.
One can also think of suppressing the  dimension-5 operator
 by imposing lepton number symmetry, broken
only at the Planck-brane. 
(This may occur, for instance, through spontaneous violation
 on the Planck-brane.)

The quantities $C_{00}$ and $A_{00}$ in the
seesaw formula (\ref{mnu}) depend on the fermion location. 
Moving the right-handed neutrino, i.e.~its former zero mode,  closer
towards the TeV-brane, we can diminish  $A_{00}$. At the
same time $C_{00}$, which also depends on the
location of $\nu$, increases. This freedom allows us to generate  
a neutrino mass of the order of $\sqrt{\Delta m^2_{\rm atm}}$,
even with a Planck-size Majorana mass as input. In order
to minimize deviations from electroweak observables, the
SM fermions, and hence the neutrinos, should be localized 
towards the Planck-brane \cite{HLS}. Taking therefore
$c_{\nu}\gsim 1/2$, the right-handed neutrino should
be localized at $0\lsim c_N \lsim 1/2$ to generate the observed
neutrino masses. In this range of parameters we have
\begin{eqnarray}
A_{00}&=&dk\left(\frac{1}{2}-c_N\right)\Omega^{2c_{N}-1}\nonumber \\
C_{00}&=&2lv_0\left(c_{\nu}-\frac{1}{2}\right)^{1/2}\left(\frac{1}{2}-c_N\right)^{1/2}
\Omega^{-c_{\nu}-1/2}
\end{eqnarray}
and the light neutrino
mass (\ref{mnu}) at leading order is given by
\begin{equation} \label{num}
m_{\nu}\approx\frac{4l^2v_0^2}{dk}\left(c_{\nu}-\frac{1}{2}\right)\Omega^{-2(c_{\nu}+c_N)}
\end{equation}
where we have used $m_M=d\cdot \delta(y)$, $H(y)=v_0\cdot \delta(y-\pi R)/\sqrt{k}$
and $\lambda_N^{(5)}=l/\sqrt{k}$.

For definiteness we take $k=M_{\rm Pl}$. To be consistent with
electroweak constraints we assume $M_{KK}=10$ TeV \cite{HLS}, which
implies $kR=10.83$. From the measured weak gauge boson
masses we find $v_0=0.043k$ \cite{H}. We take the SM
neutrino location to be $c_{\nu}=0.565$ \cite{H}. We assume a light 
neutrino mass on the order of the atmospheric mass splitting 
$\sqrt{\Delta m^2_{\rm atm}}=50$ meV.  Then eq.~(\ref{num}) leads to
a right-handed neutrino position of $c_N=0.293$. We find an effective
seesaw scale of $A_{00}=3.9\times10^{11}$ GeV and a Dirac mass
of $C_{00}=4.5$ GeV. The right-handed
neutrinos are sterile and can therefore be localized at $c<1/2$ without 
disturbing the electroweak fit. From eq.~(\ref{num})
we also observe that $m_{\nu}$ only depends on $c_{\nu}+c_N$.
The neutrino mass does not change if the left- and right-handed
neutrinos are shifted in opposite directions by the same amount. 
The lowest lying KK states consist of an almost degenerate 
pair of sterile neutrinos
with a mass of 8.5 TeV and a mass splitting of 0.1 MeV. The first
KK excitations of the SM neutrinos have a mass of 10.3 TeV
and are split by 1 MeV. The mass splittings may be affected by
radiative corrections which we have neglected in our discussion.
Clearly, the discussion can be extended to include the solar
mass splitting.

In the mass matrix (\ref{nmass}) the SM neutrinos mix 
with the left-handed KK states of the sterile neutrinos,
where the mixing angles are on the order of 
$\theta_n\approx C_{0n}/D_{N,n}$. This mixing changes the 
effective weak charge of the light neutrinos. The effective number
of neutrinos contributing to the width of the Z boson is reduced
to $n_{\rm eff}=3-\sum \sin^2\theta_n$. A similar effect occurs
if a small Dirac mass for the SM neutrinos is generated by coupling 
them to right-handed neutrinos in the bulk \cite{GN,HS3}. 
Measurements of the Z width impose the constraint 
$\delta n \lsim 0.005$ \cite{zwidth}.
For the parameter values discussed above we find
$\delta n = 2\cdot 10^{-6}$, well below the experimental sensitivity,
but still much larger than in the ordinary 4D seesaw.
The mixing is similar to the value we obtained for the model of 
ref.~\cite{HS3}. The admixture of sterile states becomes
larger if the SM neutrinos are localized closer towards the
TeV-brane, or if the KK scale is reduced. \footnote{For possibilities
to lower the KK scale see refs.~\cite{branekin,LR}.}

Similar to our discussion in ref.~\cite{HS3} the mixing between
SM neutrinos and KK sterile neutrinos considerably enhances
lepton flavor violating processes \cite{K00}, such as 
$\mu\rightarrow e\gamma$. In the warped seesaw we expect
the rates for such processes to be of the same order as
in the model of Dirac neutrino masses, which were
found to be several orders of magnitude below the
experimental bound \cite{HS3}. The branching ratio
might be brought to an experimentally interesting range if the
admixture of sterile states can be enhanced. Of course,
the setup we discussed here is crucially different from the 
model of ref.~ \cite{HS3} since the light neutrino mass
is Majorana-like. Depending on the absolute value of the
neutrino mass, this can be tested in neutrinoless double beta
decay experiments \cite{HMC}.

Finally, we briefly discuss what happens if 
the Majorana mass for the singlet neutrino is introduced
away from the Planck-brane.
If a  Majorana mass of order $M_{\rm Pl}$ is localized at the TeV-brane, 
it will be warped down to TeV-size. We expect the ``light'' neutrino mass
then to be of order GeV$^2/$TeV$\sim$ MeV. The situation is similar
if the Majorana mass is placed in the bulk. Taking it to be of order
$M_{\rm Pl}$, it completely destroys the vector-like nature of the
KK excitations, which emerge as Majorana particles with TeV-scale
masses. Again we end up with neutrino masses in the MeV-range.
Thus the warped seesaw prefers the Majorana mass to be
localized at the Planck-brane.

\section{Conclusions}
We have studied the seesaw mechanism in a warped geometry framework.
Sterile (``right-handed'') neutrinos are introduced in the bulk which 
couple to the SM neutrinos. Lepton number is broken by a Planck-size 
Majorana mass for the sterile neutrinos. If the Majorana mass is confined 
to the Planck-brane, a heavy mass scale for the seesaw is generated. 
The effective seesaw scale is of order $M_P\exp((2c-1) \pi kR)$ and 
depends on the location, i.e. 5D Dirac mass 
parameter $c$, of the sterile neutrino in the bulk. For $c<1/2$ intermediate
values of the seesaw scale emerge. For $c\approx 0.3$ light neutrinos masses
needed to explain the atmospheric and solar neutrino oscillations are obtained
without introducing any small parameters. The KK spectrum consists of
the almost degenerate excitations of the SM and sterile neutrinos, which have
masses in the TeV-range. 
It remains to be seen if the appearance of an intermediate mass scale for 
right handed neutrinos allows one to implement a successful mechanism
of leptogenesis to account for the baryon asymmetry of the universe.

\section*{Acknowledgements}
S.H.~would like to thank D.~St\"ockinger for helpful discussions.

\end{document}